\documentclass[aps,twocolumn,prc,superscriptaddress]{revtex4-1}
\usepackage[utf8]{inputenc}
\usepackage{graphicx}

\begin{document}

\title{In-medium effects for nuclear matter in the Fermi energy domain}
\author{O.~Lopez}
\affiliation{Laboratoire de Physique Corpusculaire, ENSICAEN, Université de Caen Basse Normandie, CNRS/IN2P3, F-14050 Caen cedex, France}
\author{D.~Durand}
\affiliation{Laboratoire de Physique Corpusculaire, ENSICAEN, Université de Caen Basse Normandie, CNRS/IN2P3, F-14050 Caen cedex, France}
\author{G.~Lehaut}
\affiliation{Laboratoire de Physique Corpusculaire, ENSICAEN, Université de Caen Basse Normandie, CNRS/IN2P3, F-14050 Caen cedex, France}
\author{B.~Borderie}
\affiliation{Institut de Physique Nucléaire, CNRS/IN2P3, Université Paris-Sud 11, F-91406 Orsay cedex, France}
\author{M.F.~Rivet}
\affiliation{Institut de Physique Nucléaire, CNRS/IN2P3, Université Paris-Sud 11, F-91406 Orsay cedex, France}
\author{R.~Bougault}
\affiliation{Laboratoire de Physique Corpusculaire, ENSICAEN, Université de Caen Basse Normandie, CNRS/IN2P3, F-14050 Caen cedex, France}
\author{E.~Galichet}
\affiliation{Institut de Physique Nucléaire, CNRS/IN2P3, Université Paris-Sud 11, F-91406 Orsay cedex, France}
\affiliation{Conservatoire National des Arts et Métiers, F-75141 Paris cedex 03, France}
\author{D.~Guinet}
\affiliation{Institut de Physique Nucléaire, CNRS/IN2P3, Université Claude Bernard Lyon-1, F-69622 Villeurbanne cedex, France}
\author{N.~Le~Neindre}
\affiliation{Laboratoire de Physique Corpusculaire, ENSICAEN, Université de Caen Basse Normandie, CNRS/IN2P3, F-14050 Caen cedex, France}
\author{P.~Marini}
\affiliation{Grand Accélérateur National d'Ions Lourds, CEA/DSM-CNRS/IN2P3, B.P. 5027, F-14076 Caen cedex, France}
\affiliation{Centre d’Études Nucléaires de Bordeaux-Gradignan, CNRS/IN2P3 - Université de Bordeaux I, F-33175 Gradignan cedex, France}
\author{P.~Napolitani}
\affiliation{Institut de Physique Nucléaire, CNRS/IN2P3, Université Paris-Sud 11, F-91406 Orsay cedex, France}
\author{M.~Pârlog}
\affiliation{Laboratoire de Physique Corpusculaire, ENSICAEN, Université de Caen Basse Normandie, CNRS/IN2P3, F-14050 Caen cedex, France}
\author{E.~Rosato}
\affiliation{Dipartimento di Scienze Fisiche e Sezione INFN, Università di Napoli "Federico II", I-80126 Napoli, Italy}
\author{G.~Spadaccini}
\affiliation{Dipartimento di Scienze Fisiche e Sezione INFN, Università di Napoli "Federico II", I-80126 Napoli, Italy}
\author{E.~Vient}
\affiliation{Laboratoire de Physique Corpusculaire, ENSICAEN, Université de Caen Basse Normandie, CNRS/IN2P3, F-14050 Caen cedex, France}
\author{M.~Vigilante}
\affiliation{Dipartimento di Scienze Fisiche e Sezione INFN, Università di Napoli "Federico II", I-80126 Napoli, Italy}
\collaboration{INDRA collaboration}

\begin{abstract}
We study nuclear stopping in central collisions for heavy-ion induced reactions 
in the Fermi energy domain,  between $15$ and $100$ A\,\textrm{MeV}. Using the large 
dataset of exclusive measurements provided by the $4\pi$ array \emph{INDRA}, we 
determine the relative degree of stopping as a function of system mass and 
bombarding energy. We show that the stopping can be directly related to the 
transport properties in the nuclear medium. By looking specifically at free 
nucleons (here protons), we present for the first time a comprehensive body of 
experimental results concerning the mean free path, the nucleon-nucleon cross-section and in-medium effects 
in nuclear matter. It is shown that the mean free 
path exhibits a maximum at $\lambda_{NN}=9.5 \pm 2$ \textrm{fm}, around  $E_{inc}=35-40$ A\,\textrm{MeV} incident energy 
and decreases toward an asymptotic value 
$\lambda_{NN}= 4.5 \pm 1$ \textrm{fm} at $E_{inc} = 100$ A\,\textrm{MeV}. After accounting for Pauli blocking 
of elastic nucleon-nucleon collisions, it is shown that the effective in-medium 
\emph{NN} cross section is further reduced compared to the free value in this energy 
range. Therefore, in-medium effects cannot be neglected in the Fermi energy 
range. These results bring new fundamental inputs for microscopic 
descriptions of nuclear reactions in the Fermi energy domain.
\end{abstract}

\maketitle

\section{Introduction}

Transport properties in nuclear matter contribute to the determination of the 
equation of state via the underlying in-medium properties of the nuclear 
interaction and are one of the fundamental ingredients for microscopic models 
\cite{Onishi1995}-\cite{Fuchs2006}. They are also critical in the description 
of the supernova core collapse and the formation of a neutron star \cite{Lattimer2004}. 
These properties can be probed with the help of heavy-ion induced collisions (\emph{HIC}) by looking at 
dissipation phenomena in terms of energy and isospin transport and thus related to the stopping. In the Fermi 
energy domain, transport features 
should exhibit the interplay between mean-field (nuclear degrees of 
freedom) and individual (nucleonic degrees of freedom) effects, especially when 
looking at 
the energy dissipation reached in central collisions \cite{Lehaut2010}. 

From a theoretical point of view, the knowledge of the dissipation mechanism 
for nuclear matter in HIC is related to the properties of the mean-field itself 
via the 
1-body dissipation (nuclear friction and viscosity) and nucleon-nucleon (\emph{NN}) 
collisions via 2-body 
dissipation in the nuclear medium. In the studied energy range, below $100$ A\,\textrm{MeV}, 
only \emph{NN} elastic channels must be considered. At low incident energy \emph{i.e.} lower than the 
Fermi energy, where mean-field effects prevail, \emph{NN} collisions are strongly 
suppressed due to the fermionic nature of nucleons, known as Pauli 
blocking. At high incident energy, while the available space for 
\emph{NN} collisions increases, the situation is 
the opposite since the mean-field becomes less and less attractive while \emph{NN} 
collisions become important 
\cite{Durand2006,Cugnon1996,Lukasik2005}. Then, one expects the in-medium \emph{NN} 
cross section to be very small at low incident energy and to become sizeable (asymptotically approaching the \emph{free} \emph{NN} cross section) as the incident 
energy becomes significantly higher than the Fermi energy. In this framework, 
one may wonder what is the magnitude of in-medium effects induced by many-body 
correlations in nuclear reactions. Numerous theoretical approaches show that 
the cross section has to be properly renormalized in order to account for the \emph{effective} \emph{NN} collision rate in \emph{HIC} \cite{LiMachleidt1993} and 
several effects must be considered. First, the Pauli blocking effect 
discussed above reduces the \emph{NN} collision rate \cite{Durand2006} and can be 
viewed as a ``trivial'' two-body correlation due the quantal nature of protons 
and neutrons. Higher-order correlations in \emph{NN} collisions due to the high-
density conditions encountered in central collisions \cite{LiMachleidt1993,
Cugnon1996,Durand2006} can also come into play as shown in some theoretical 
works done in the past two decades \cite{LiMachleidt1993,Frick2005,Soma2008,
Rios2012}. They mainly use relativistic mean-field approaches with realistic (effective) nucleon-nucleon interactions. They conclude that 
the in-medium \emph{NN} cross sections are isopin-dependent, and have to be renormalized (reduced) 
in the nuclear medium as compared to free vacuum values. They 
show that the nucleon mean free path is large, typically larger than the 
nucleus size for $E_{inc}/A \leq 100$ \textrm{MeV}, and decreases toward a saturation 
value $\lambda_{NN}=4-5 ~fm$ at high incident energy, for $E_{inc}/A \geq 100$ 
 \textrm{MeV} \cite{Rios2012}. Thus, the situation at high energy, where the mean free 
path is supposed to be almost constant, is quite clear. This is not the case in 
the Fermi energy domain; indeed, in-medium effects and especially quenching 
factors for the \emph{NN} cross section are largely unknown in the range $E_{inc}/A=10-
100$ \textrm{MeV} \cite{Westfall1993} and have to be constrained experimentally. 

From an experimental point of view, nuclear stopping has been determined by the \emph{FOPI} collaboration 
for the \textrm{Au}+\textrm{Au} system in the incident energy range $90-1930$ A\,\textrm{MeV} using 
several 
observables \cite{Andronic2006}. The study concluded that there is a broad 
plateau of maximal stopping between $200$ and $800$ A\,\textrm{MeV} \cite{Andronic2006}. For 
the Fermi energy domain, the situation is quite different. In our previous 
paper \cite{Lehaut2010}, we have shown that the stopping measured as the ratio 
between transverse and longitudinal energies of the reaction products can probe the energy dissipation 
in central collisions and shed light on the dissipation mechanism itself. A 
transition was observed from a 1-body to a 2-body dissipation mechanism as a 
function of the incident energy. The transition occurs around $35$ A\,\textrm{MeV} (close 
to the Fermi energy at saturation density), whatever the system size. It 
corresponds to a \emph{minimal} value for the stopping \cite{Lehaut2010}. These 
results call for an extended analysis of the experimental data. The purpose of 
this paper is then a continuation of this work, where we try to relate the observed 
dependence for the stopping in central collisions to the nucleon mean free path $\lambda_{NN}$ and 
the corresponding cross section $\sigma_{NN}$ in the nuclear medium.  

\section{Experimental considerations}

\subsection{$INDRA$ dataset}

In this analysis, we use the full \emph{INDRA} dataset for symmetric or nearly-
symmetric systems recorded along the past two decades at \emph{GANIL} and \emph{GSI} facilities. 
The experimental data are exclusive and corresponds to a nearly complete 
detection of all charged products of the reaction thanks to the powerful \emph{INDRA} $4\pi$ array \cite{Pouthas1995}. Details concerning the data collection 
can be found in \cite{Borderie2002,Plagnol1999,Metivier2000,Hudan2003,Lukasik2002,LeFevre2004}. 
Table I shows some basic characteristics of the studied systems.

\begin{table}[h*]
\begin{tabular}{|c|c|c|c|c|c|c|}
\hline
System & $A_{tot}$ (amu) & $E_{inc}$ (A\,\textrm{MeV}) & $Asym$  & $(N/Z)_{tot}$ \\
\hline
$^{36}Ar+KCl$ & 72 & 32-74 (5) & 0 & 1 \\
$^{36}Ar+^{58}Ni$ & 94 & 32-95 (7) & 0.23 & 1.04 \\
$^{58}Ni+^{58}Ni$ & 116 & 32-90 (7) & 0 & 1.07 \\
$^{129}Xe+^{129}Sn$ & 248 & 15-100 (14) & 0.04 & 1.39 \\
$^{181}Ta+^{197}Au$ & 378 & 33,40 (2) & 0.04 & 1.49 \\
$^{197}Au+^{197}Au$ & 394 & 40-100 (4) & 0 & 1.49 \\
$^{155}Gd+^{238}U$ & 393 & 36 (1) & 0.21 & 1.59 \\
$^{208}Pb+^{197}Au$ & 405 & 29 (1) & 0.03 & 1.52 \\
$^{238}U+^{238}U$ & 476 & 24 (1) & 0 & 1.59 \\

\hline
\end{tabular} 
\label{INDRA_dataset}
\caption{Characteristics of the $42$ (quasi-)symmetric systems measured with 
\emph{INDRA} and analysed in this study. The mass asymmetry in the entrance channel 
is defined as: $Asym=|A_{projectile}-A_{target}|/(A_{projectile}+A_{target})$. 
Numbers in brackets in the third column indicate the number of measured bombarding 
energy. }
\end{table}

We note that the data cover a broad domain of incident energy, here from $15$ 
up to $100$ A\,\textrm{MeV}, and concern $42$ systems with a total mass between $72$ and 
$476$ mass units. Isospin is here comprised between $N=Z$ and $N/Z\approx1.6$. 
This constitutes, to our knowledge, the largest body of experimental data in 
the Fermi energy domain covered with the same setup. Two systems in Table I are 
not fully symmetric ($^{36}$\textrm{Ar}$+^{58}$\textrm{Ni} and $^{155}$\textrm{Gd}$+^{238}$\textrm{U}) but still 
present a small mass asymmetry; they have been taken in order to cover more 
efficiently the mass/energy domain of the analysis. In the 
following, we will display all quantities as a function of the incident energy 
in the laboratory frame. In 
a more general perspective, one should prefer to use the available center-of-mass energy, especially when including asymmetric systems. 

\subsection{Event selection}

\begin{figure}[h*]
 \begin{center}
 \includegraphics[width=0.5\textwidth]{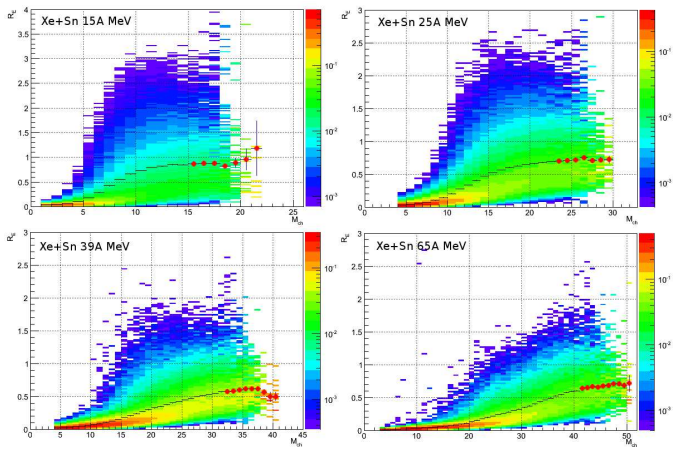}
 \caption{Isotropy ratio $R_E$ as a function of total charged multiplicity $M_{ch}$ for 
 \textrm{Xe}+\textrm{Sn} at $15$, $25$, $39$ and $65$ A\,\textrm{MeV}. The black histograms show the mean values and the grey (red) symbols 
indicate the event selection. All $M_{ch}$ bins have been normalized 
to the same number of entries (color online).}
 \label{RE_Mch}
\end{center}
\end{figure}

In the following, we want to probe the degree of stopping in central collisions.
 We have then chosen to study the very dissipative collisions, that produce the 
highest charged particle multiplicities $M_{ch}$; we use the multiplicity 
selection as a \emph{minimum bias} selector. By doing so, we minimize the 
inevitable auto-correlations between the event selection and the observable of 
interest, here the isotropy ratio built upon the kinematical properties of 
particles. We use a scalar variable -$M_{ch}$- as event selector in order to 
look at a vector observable, namely the energy isotropy ratio $R_E$. This 
latter is defined on an event-by-event basis: 

\begin{equation}
 R_E = \frac{1}{2}\frac{\Sigma_{i}^{N} E_i^{\perp}}{\Sigma_{i}^{N} E_i^{//}} 
 \label{RE}
\end{equation}

where $E_i^{\perp}$ and $E_i^{//}$ are the transverse and longitudinal center-of-mass (c.m.) energies for particle $i$. The 
summation is done over the total number $N$ of (detected) reaction products in 
the selected event. By construction, $R_E$ is equal to $1$ for an isotropic 
emission, $<1$ for an elongated emission along the longitudinal direction given 
by the beam direction and $>1$ for preferential emission in the plane transverse
 to the beam direction. Since we are looking at \emph{INDRA} data, the sum is 
restricted to charged products only, but we however benefit from the excellent 
$4\pi$ coverage of the experimental apparatus. Fig. \ref{RE_Mch} presents the 
correlation between the charged particle multiplicity $M_{ch}$ and the isotropy 
ratio $R_E$ obtained from \emph{INDRA} data for the four \textrm{Xe}+\textrm{Sn} systems at 
$15,25,39$ and $65$ A\,\textrm{MeV}. The \emph{INDRA} trigger was set to $M_{ch}>3$, 
allowing to record $60-80\%$ of the total reaction cross section \cite{Lehaut2010}. The $M_{ch}$ 
bins have been normalized to the same number of entries in order to reduce the 
statistical fluctuations as done in \cite{Lehaut2010}. The black histograms are the 
corresponding mean $R_E$ values.

This correlation presents a saturation at the highest multiplicity values, 
represented by the black histogram in Fig. \ref{RE_Mch}. We then define a 
multiplicity cut in order to retain the events corresponding to $M_{ch}>M_{ch}^{cut}$, 
visible as the symbols onto Fig. \ref{RE_Mch}. The multiplicity cut 
clearly depends on the system and has been set using the same strategy for the 
whole dataset of Table I. The selection retains typically between $50$ and 
$150$ \textrm{mb}, thus corresponding, assuming that only the most central collisions 
are selected, to an impact parameter range between $0$ and $b=1-1.5$ \textrm{fm} (\emph{i.e.} 
$1-2\%$ of the detected events). Alternatively, we could have used a fixed 
value of the total cross section for all systems (for example the lowest one: 
$50$ \textrm{mb}), but this would not change substantially the results concerning the 
extracted $R_E$ values.

\subsection{Particle selection}

In order to probe the nucleon properties in nuclear medium, we have to focus 
specifically on free nucleons. 
They indeed carry genuine information about \emph{NN} collisions, \emph{i.e.} out of 
any coalescence phase nor clusterization into fragments occuring during the course 
of the collision \cite{Zhang2011}. 

\begin{figure}[h*]
 \begin{center}
\includegraphics[width=0.5\textwidth]{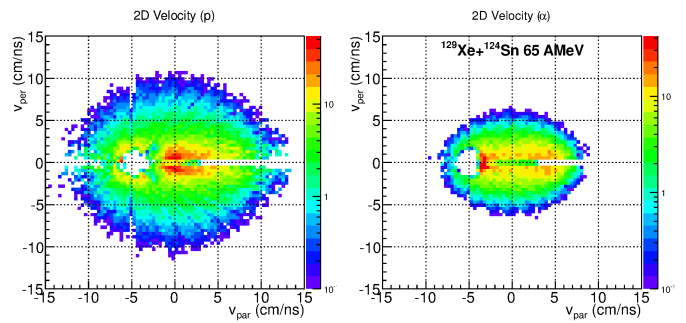} 
\caption{\emph{c.m.} transverse versus parallel velocities for protons (left) and $\alpha$ 
particles (right) in invariant cross section, for the selected central events of the $^{129}$\textrm{Xe}+$^{124}$\textrm{Sn} system at $65$ A\,\textrm{MeV} (color online).}
\label{Vper_vpar_Z=1}
\end{center}
\end{figure}

In fig. \ref{Vper_vpar_Z=1}, we can see the \emph{c.m.} velocity plots in invariant 
cross section for protons (left) and $\alpha$ particles (right), for the 
selected central events for the $^{129}$\textrm{Xe}$+^{124}$\textrm{Sn} system at $65$ A\,\textrm{MeV}. Protons clearly exhibit different kinematical features compared to $\alpha$ particles, with a strong emission located at mid-rapidity and an extension to high 
transverse velocities suggesting a non-equilibrium emission. We will then consider that protons are predominantly produced before thermalization of the produced hot nuclei and not from secondary decay as already seen in a previous study for \textrm{Xe}+\textrm{Sn} central collisions \cite{Hudan2003}. 

In the following, we limit our study to protons, for which we 
compute the isotropy ratio $R_E$ and call it hereafter $R^{p}_E$. To avoid the statistical 
fluctuations coming from the event-by-event determination of $R^{p}_E$, we 
rather compute the isotropy ratio from the \emph{full} set of protons selected 
by the multiplicity cut, considering thus all protons detected in these events 
as though they are coming from a single event. It is worthwhile to note that 
this procedure weakly lowers ($5-10\%$) the mean values for $R^{p}_E$ as compared to the event-by-event determination.

\section{Stopping in nuclear matter}

\subsection{Stopping ratio for protons}

\begin{figure}[h*]
 \begin{center}
\includegraphics[width=0.5\textwidth]{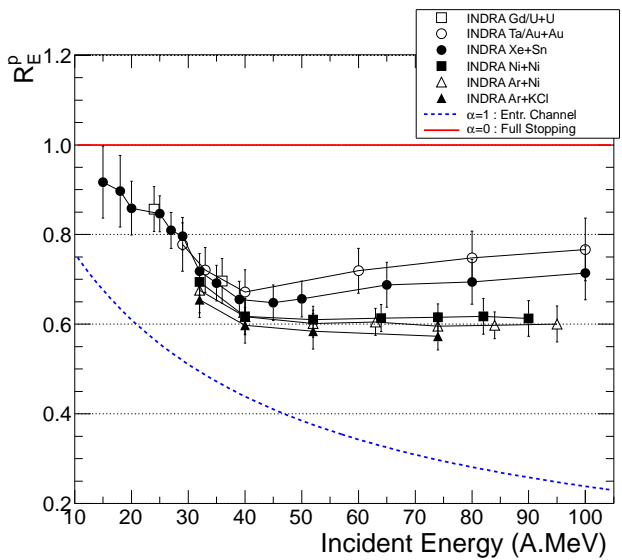}
 \caption{Mean isotropy ratio for protons $R^{p}_E$ as a function of  
energy. The symbols represent the different studied systems. The lower dashed curve represents the expected value for the entrance channel (no stopping) and the upper straight line for the full stopping (color online).}
 \label{RE_Z=1}
\end{center}
\end{figure}

Applying the protocol presented in the previous section, we compute the isotropy
 ratio $R^{p}_E$ for the $42$ different systems listed in Table I. The results are presented in 
Fig. \ref{RE_Z=1} as a function of the incident energy.

The error bars in Fig. \ref{RE_Z=1} correspond to the statistical errors 
supplemented by an estimate of the systematic errors coming from the experimental determination for $R^{p}_{E}$. For these latter, we use the same prescription 
as in \cite{Lehaut2010}; we consider here a ``reasonable'' variation for the 
multiplicity cuts $M_{ch}^{cut}\pm 1$) in the event selection and take 
the corresponding $R^{p}_E$ intervals as an estimator of the systematic 
errors. They are found to contribute for more than half of the total error bars,
 depending on systems. In the following, the error bars displayed on all 
computed quantities will derive from these ones, thus will incorporate not only 
statistical but also (some) systematic errors.

We now compare the results for $R^{p}_E$ to the ones obtained for $R_E$ in \cite{Lehaut2010} 
for all particles (no proton selection). Although the values 
are systematically higher, we find a similar behavior; a quite steep decrease 
from low incident energy to Fermi energy followed by a flattening or even a 
modest increase for the isotropy ratio at higher incident energies. We also get 
a similar mass scaling for the different systems; the heavier the system, the 
higher the isotropy ratio is. This supports the fact that the stopping, \emph{i.e.} the conversion from longitudinal to transverse energy, is related to the 
number of participants in the system as in a \emph{Glauber} description of the collision \cite{Charagi1990}. 
The difference with \cite{Lehaut2010} comes from the 
location in incident energy of the transition between the 2 
regimes; it is rather $30-35$ A\,\textrm{MeV} in \cite{Lehaut2010} while it is slightly 
above in the present study, between $35$ and $40$ A\,\textrm{MeV}. We can also notice that 
the mean isotropy ratio is always below $1$ and thus on average the proton 
momentum distribution never achieves the isotropy which we associate with full 
stopping for the selected events. This was also the case in \cite{Lehaut2010}. 

To get more quantitative values for the stopping, the isotropy ratio is compared
 to two extreme values. They are computed by assuming two \emph{Fermi} spheres 
in $p$-space separated by the relative momentum corresponding to  $E_{inc}$, 
which is the incident energy and $\alpha$ a parameter equal to $1$ 
for complete transparency (no dissipation, lower dashed curve in blue) and $0$ for full stopping
 (upper straight line in red). A straightforward calculation for the isotropy ratio $R_E(\alpha)$ can be obtained analytically:

\begin{equation}
R_E(\alpha)=\frac{1}{1+5 \alpha x/4}
\label{REdiss}
\end{equation}

where $x=E_{inc}/E_{Fermi}$ and $E_{Fermi}=38$ \textrm{MeV} is the Fermi energy at saturation density. An estimate 
for the stopping reached in our dataset of central events is then given by the 
normalized quantity $\mathcal{S}$, called hereafter stopping ratio, such as: 

\begin{equation}
\mathcal{S}=\frac{R^{p}_E-R_E(\alpha=1)}{R_E(\alpha=0)-R_E(\alpha=1)}
\label{distance}
\end{equation}

This quantity is always positive since $R^{p}_E > R_E(\alpha=1)$ and $R_E(\alpha=0)> R_E(\alpha=1)$.

\begin{figure}[h*]
 \begin{center}
\includegraphics[width=0.5\textwidth]{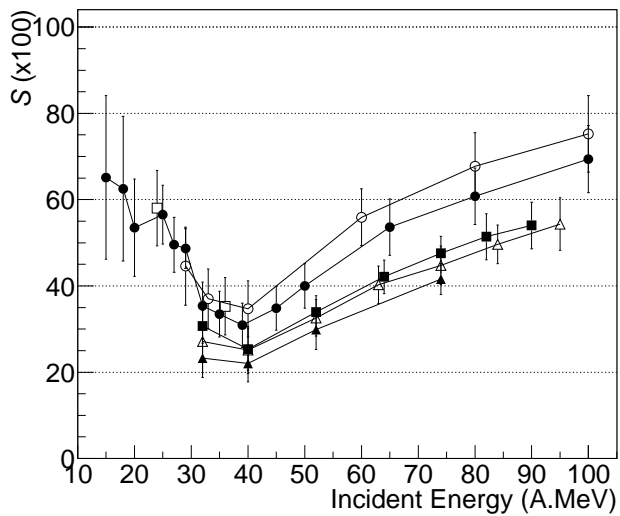}
\caption{Stopping ratio $\mathcal{S}$ (see text) as a function of incident 
energy. Symbols are the same as for Fig. \ref{RE_Z=1}.}
 \label{Stopping_Z=1}
\end{center}
\end{figure}

Fig. \ref{Stopping_Z=1} displays the stopping ratio $\mathcal{S}$ (in percentage) and 
emphasizes the location of the minimal stopping value around $40$ A\,\textrm{MeV} 
altogether with the 2 different regimes for dissipation. As expected from Fig. \ref{RE_Z=1}, 
we find stopping ratio values ranging between $0$ and $1$. The 
minimum stopping,  $\mathcal{S}_{min}=20-35 \%$, depends on the system. It is 
larger for the heavier systems ($\mathcal{S}_{min}\approx35\%$), here \textrm{Ta}/\textrm{Au}+\textrm{Au} 
and \textrm{U}+\textrm{U}. Let us recall that $R^p_{E}$ is larger than $R_{E}$. We 
attribute this effect to the fact that free nucleons, supposedly \emph{NN} collisions, have to be emitted \emph{outside} the two nuclei in $p$-space. We are going to develop this point in the next section.

\subsection{Stopping ratio and \emph{NN} collisions}

\begin{figure}[h*]

\includegraphics[width=0.5\textwidth]{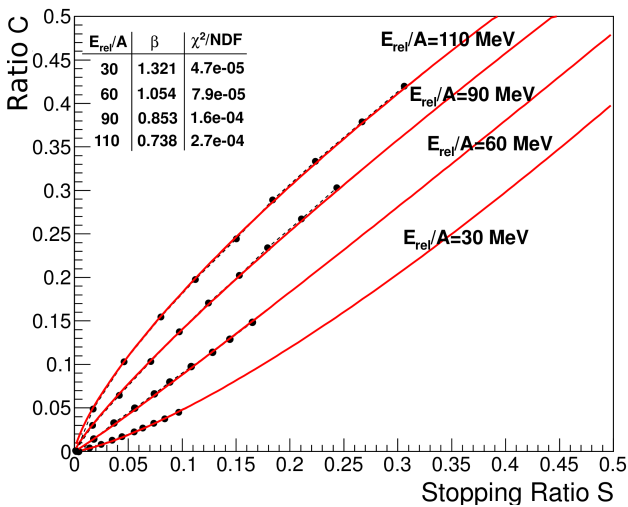}
\caption{Correlation between the 
probability $\mathcal{C}$ of accepted collisions and the stopping ratio $\mathcal{S}$ for two Fermi spheres separated by a relative energy of $30,60,90,110$ A\,\textrm{MeV} (symbols). The curves correspond to the fits $S^{\beta}$, with the $\beta$ values displayed in the inserted table and their corresponding reduced $\chi^2$ values.}
\label{Fermi_spheres}
\end{figure}

In this section, we want to link the stopping ratio $\mathcal{S}$ to a quantity 
related to the amount of \emph{NN} collisions, taking into account the proper available phase space in momentum. 
To do so, we use a simple Monte-Carlo 
simulation, taking again the two Fermi spheres described in the previous 
section, and we implement elastic \emph{NN} collisions in a semi-classical way; to perform such collisions, we 
pick randomly one nucleon from the projectile and one from the target and 
rotate the corresponding momenta around their own \emph{c.m.} frame. In this procedure, we do not consider multiple scatterings nor the Pauli blocking in the final state between scattered nucleons. We accept the collision if this motion brings the two nucleons outside the 
two Fermi spheres, according to a probability corresponding to the level of $NN$ collisions which are supposed to be really produced, in order to mimick the fact that in-medium effects can affect the number of allowed $NN$ collisions. For each incident energy, we perform a run of $100,000$ collisions in order to scan extensively the corresponding available phase space, and we vary the probability between 0 (no allowed collision) and 1 (fully allowed collisions given the available phase space) for each run. We then register the ratio $\mathcal{C}$ between accepted and attempted collisions for each value of the probability at a given incident energy. The obtained correlations between $\mathcal{C}$ and the stopping ratio $\mathcal{S}$ is displayed in Fig. \ref{Fermi_spheres} for incident energies between $30$ and $110$ A\,\textrm{MeV}, each point corresponding to one probability value for a given incident energy. 

This procedure accounts for the Pauli exclusion principle in $p$-space and 
allows to compute the isotropy ratio $R_E$, (Eq. \ref{RE}) and the stopping 
ratio $\mathcal{S}$ (Eq. \ref{distance}). By varying the incident energy between $30$ and $110$ A\,\textrm{MeV}, we find that $\mathcal{S}$ is related to the ratio of accepted \emph{NN} collisions $\mathcal{C}$ by the following empirical formula as illustrated by the curves in Fig. \ref{Fermi_spheres}: 

\begin{equation}
\mathcal{C} \approx \mathcal{S}^{\beta(E_{inc})}
\label{collision}
\end{equation}
 
with $\beta=1.32$ at $E_{inc}=30A$ \textrm{MeV}, and $\beta=0.74$ at $E_{inc}=110A$ \textrm{MeV}. The quality of this approximation is illustrated by the agreement between the fits (curves) and the symbols in Fig. \ref{Fermi_spheres}. This is also quantified by the reduced $\chi^2$ values in the inserted table. The energy dependence for $\beta$ is then simply parametrized as a smooth quadratic dependence upon the incident energy in A\,\textrm{MeV}: $\beta(E_{inc})=1.643-1.155.10^{-2}E_{inc}+2.974.10^{-5}E^2_{inc}$. 
This parametrization nicely describes the correlation between $\mathcal{C}$ and $\mathcal{S}$ for the considered energy range with a good level of accuracy; it can be seen as the functional form between the stopping ratio $\mathcal{S}$ and the percentage of \emph{NN} collisions $\mathcal{C}$ for the corresponding available phase space. In the following, we will use this quantity $\mathcal{C}$ calculated from Eq. \ref{collision} to extract information on $NN$ collisions. 

\subsection{Mass scaling and characteristic length}

\begin{figure}[h*]
 \begin{center}
\includegraphics[width=0.5\textwidth]{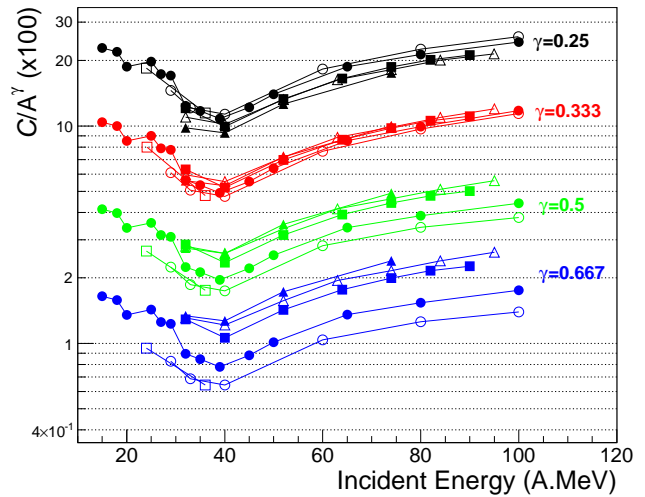}
\caption{Scaled quantity $\mathcal{C}/A_{tot}^{\gamma}$ (see text), here multiplied by 100, as a function
 of incident energy for $\gamma=0.25,0.333,0.5,0.667$. Symbols are the same as for Fig. \ref{RE_Z=1} (color online).}
 \label{RE_scaled_Z=1}
\end{center}
\end{figure}

To understand the mass hierarchy observed in Figs. \ref{RE_Z=1}-\ref{Stopping_Z=1}, we scale the latter quantity $\mathcal{C}$ by $A_{tot}^{\gamma}$, $A_{tot}$ being the total mass number of the system, and $\gamma$ varying between $1/4-2/3$ . The results are shown in Fig. \ref{RE_scaled_Z=1}. For $\gamma\approx\frac{1}{3}$, all experimental points collapse on a single curve for the whole range of incident energy and for all systems; the agreement is somehow particularly impressive for incident energies above the Fermi energy.

This result suggests to define a characteristic quantity $A_{tot}^{1/3}$, homogeneous 
to a length, connected to the radial extent of the system formed in central collisions. This length appears to be a key quantity for describing the amount of stopping and hence the percentage of \emph{NN} collisions. In a \emph{Glauber} picture, this can be seen as the \emph{characteristic length} associated to \emph{NN} collisions in nuclear matter. From this, we can infer that the corresponding reduced value $\mathcal{C}/A_{tot}^{1/3}$ is related to the associated mean free path for \emph{NN} collisions. 

\section{In-medium effects}

\subsection{Nucleon mean free path}

In this section, we estimate the mean free path for a nucleon from the stopping ratio $\mathcal{S}$ and the related quantity $\mathcal{C}$. We postulate from the previous findings that the mean free path $\lambda_{NN}$ can be simply expressed as the \emph{inverse} of $\mathcal{C}$: 

\begin{equation}
 \lambda_{NN} \approx L/\mathcal{C}
 \label{lambdaNN}
\end{equation}

where $L$ is a characteristic length proportional to $A_{tot}^{1/3}$, taken equal to the 
average nuclear radius $L=r_0 A^{1/3}$ with $r_0=1.2$ \textrm{fm} and $A=A_{tot}/2\approx A_{projectile}\approx A_{target}$. $L$ can be interpreted as a quantity related to the average distance travelled by a nucleon. Also, we assume implicitly that the quantity $\mathcal{C}=S^{\beta}$ corresponds to the percentage of \emph{NN} collisions when the two incoming nuclei fully overlap in $r$-space as one can expect for central collisions. At this stage, we do not expect any significant change for $\lambda_{NN}$ if we consider a higher density ($\rho/\rho_0 \approx 1.2$), hence a slightly smaller $L$ value, for the colliding system.
 
\begin{figure}[h*]
 \begin{center}
\includegraphics[width=0.5\textwidth]{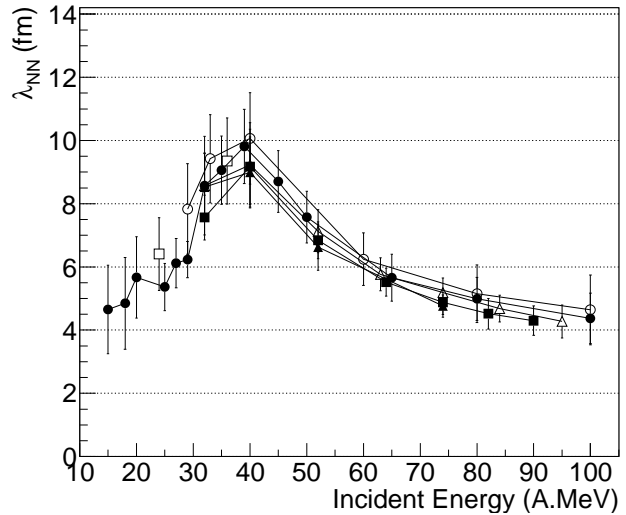}
\caption{Mean free path for a nucleon in nuclear matter as a function of incident energy. Symbols are the same as for Fig. \ref{RE_Z=1}.}
 \label{MFP_Z=1}
\end{center}
\end{figure}

Applying Eq. \ref{lambdaNN}, we plot the results in Fig. \ref{MFP_Z=1}. We see that $\lambda_{NN}$ is maximum around $E_{inc}=35-40$ A\,\textrm{MeV}, thus corresponding to a minimum value for 
the stopping as observed in Figs. \ref{RE_Z=1}-\ref{Stopping_Z=1}, and reaches $\lambda_{NN} = 9.5~\pm 2$ \textrm{fm}. This depicts the fact that the Pauli principle suppresses to a large extent \emph{NN} collisions at low incident energy 
and consequently increases the mean free path around the Fermi energy \cite{Durand2006}. 
The decrease observed at lower incident energy is here attributed 
to mean-field effects, for which the dissipation mechanism is mainly provided 
by 1-body rather than 2-body dissipation. In this energy domain, the stopping 
ratio $\mathcal{S}$ (and consequently $\mathcal{C}$) should be certainly computed in a more appropriate way since the sudden approximation taken here as a reference for $\alpha=1$ (no mean-field dissipation, see eq. \ref{REdiss}) should be less valid. This will be extensively studied in a forthcoming paper. 

If we now focus on the high energy domain, \emph{i.e.} above the Fermi energy, 
we note a continuous decrease of $\lambda_{NN}$, whatever the system size, 
toward an asymptotic value corresponding to $\lambda_{NN}=4.5\pm 1~fm$ above $100$ A\,\textrm{MeV}. These values are compatible with both experimental data \cite{Renberg1972,Nadasen1981} and recent theoretical studies \cite{Rios2012} around and above 
$100$ A\,\textrm{MeV}. This agreement also suggests that the characteristic length 
$L$ is indeed closely related to the nuclear radius of the colliding system and justifies \emph{a posteriori} our assumption.

\subsection{Nucleon-nucleon cross section}

\begin{figure}[h*]
 \begin{center}
\includegraphics[width=0.5\textwidth]{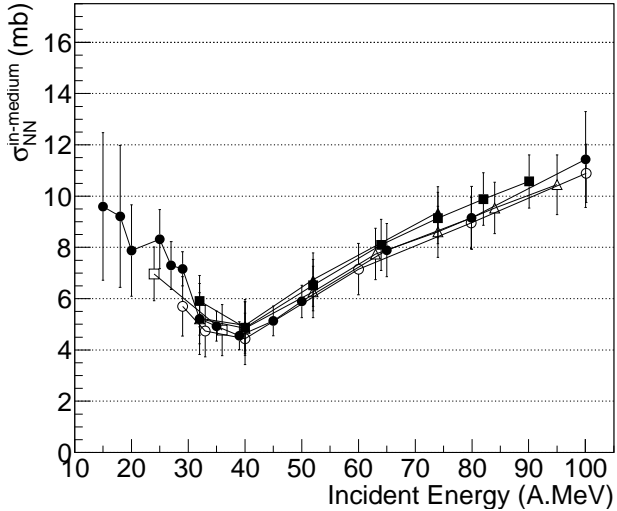} 
\caption{In-medium nucleon-nucleon cross section as a function of incident 
energy for $\rho=1.2\rho_0$ (see text). Symbols are the same as for Fig. \ref{RE_Z=1}.}
 \label{SNN_Z=1}
\end{center}
\end{figure}

From our estimated mean free paths, we can now determine the in-medium nucleon-nucleon 
cross section by taking the standard formula from kinetic theory: $\sigma_{NN} \approx 1/(\rho \lambda_{NN})$. 
We choose here the density $\rho=1.2\rho_0$ with $\rho_0=0.17$ \textrm{fm}$^{-3}$ since we are looking at central collisions where the two nuclei are supposed to strongly overlap in $r-$space. The density value taken here is considered as a standard value concerning the incident energy range $30-100$ A\,\textrm{MeV} \cite{Danielewicz1995}. We could have taken a more sophisticated energy-density dependence, but it would not affect the results as explained later on. We then obtain the values of $\sigma_{NN}$ displayed in Fig. \ref{SNN_Z=1}, with an asymptotic value at high energy close to $12$ \textrm{mb}. In the following, we will compare these extracted in-medium cross-sections to the free values in vacuum. 

\subsection{In-medium effects}

\begin{figure}[h*]
 \begin{center}
\includegraphics[width=0.5\textwidth]{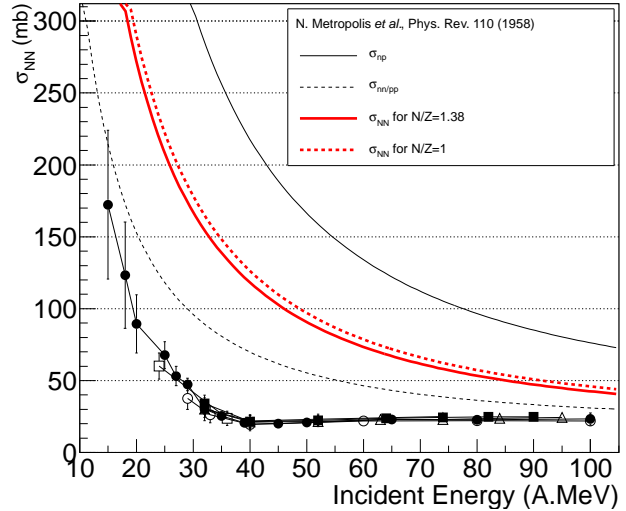} 
\caption{In-medium nucleon-nucleon cross section corrected from Pauli effects (see text) as a function of incident energy. The different curves correspond to the free values for $nn/pp$, $np$ and \emph{NN} cross section for two isospin ratios: $N/Z=1,1.38$. Symbols are the same as for Fig. \ref{RE_Z=1}.}
 \label{Free_SNN_Z=1}
\end{center}
\end{figure}

To disentangle the different in-medium effects, we start by evaluating the Pauli blocking. Several methods can be employed \cite{Chen2013,Su2013}. 
We use in this study the simple prescription of Kikuchi and Kawai \cite{Kikuchi1968} where the probability $P$ to perform a \emph{NN} collision is given by : 

\begin{eqnarray*}
P(\zeta) &= 1-7\zeta/5~~~~\mathrm{for}~\zeta\leq0.5~~~~~~~~~~~~~~~~~~~~~~~~~\\
P(\zeta) &= 1-7\zeta/5+2\zeta(2-1/\zeta)^{5/2}/5~~~~\mathrm{for}~\zeta>0.5
\end{eqnarray*}

with $\zeta = E_{Fermi}/(E_{inc}+E_{Fermi})$. $E_{inc}$ is the incident energy between the two incoming nuclei, and $E_{Fermi}=38(\rho/\rho_0)^{2/3}$ is the Fermi energy for a nucleus at density $\rho$. By dividing $\sigma_{NN}$ as reported in Fig. \ref{SNN_Z=1} by $P(\zeta)$, we thus obtain nucleon-nucleon cross sections, out of Pauli effects, which have to be compared to the standard free values \cite{Metropolis1958} as shown by Fig. \ref{Free_SNN_Z=1}.

The curves correspond to the values for neutron-neutron (\emph{nn})/proton-proton (\emph{pp}), neutron-proton (\emph{np}) and a combination of both \cite{Charagi1990} to 
get $\sigma_{NN}$ for a given nucleus with $N/Z=1,1.38$ ($N/Z$ values corresponding to some of the systems studied here, see Table~I). We observe that the experimental in-medium \emph{NN} cross sections are systematically lower than the free cross sections, tending however to recover the free values at high incident energy, well 
above $100$ A\,\textrm{MeV}. This shows that additional in-medium effects, outside Pauli effects, are indeed present and have to be taken into account for renormalizing the 
free nucleon-nucleon cross sections in nuclear matter. Note however that we may 
have underestimated the Pauli blocking as it was determined only in $p$-space instead of the full phase space \cite{Chen2013}. 

\begin{figure}[h*]
 \begin{center}
\includegraphics[width=0.5\textwidth]{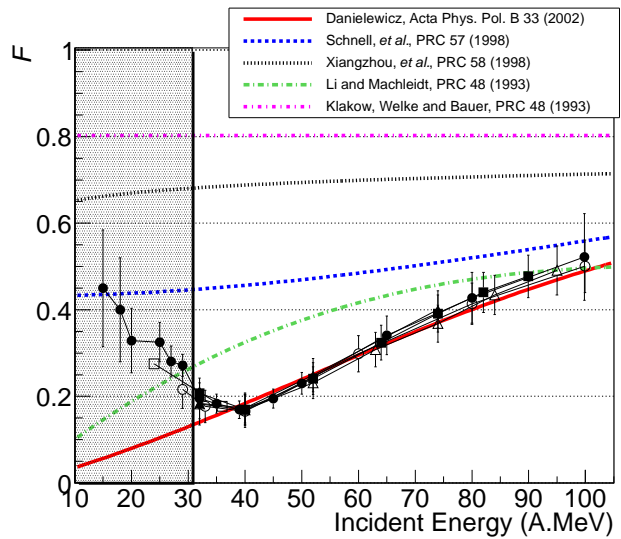} 
\caption{In-medium factor $\mathcal{F}=\sigma_{NN}^{in-medium} / \sigma_{NN}^{free}$ for the nucleon-nucleon cross section in nuclear matter. The different curves correspond to some parametrizations used in transport models. Symbols are the same as for Fig. \ref{RE_Z=1}.}
 \label{Factor_Z=1}
\end{center}
\end{figure}
 
To get more quantitative results, we compute the in-medium factor 
$\mathcal{F}=\sigma_{NN}^{in-medium} / \sigma_{NN}^{free}$ from Fig. \ref{Free_SNN_Z=1}. It is displayed in Fig. \ref{Factor_Z=1} where we restrict our discussion to the incident energy range $30-100$ A\,\textrm{MeV}. As already discussed, the data 
could give nonphysical results at incident energy lower than the Fermi 
energy (shaded area in Fig. \ref{Factor_Z=1}) and should not be taken into 
account in this comparison. For this specific energy range, we should apply a 
special treatment to the stopping ratio $\mathcal{S}$, with mean-field effects 
properly evaluated, in order to be compared with theoretical prescriptions. 
These aspects will be studied in a forthcoming paper as already mentioned. 

The reduction factor $\mathcal{F}$ strongly evolves with incident energy, between
$0.2$ and $0.5$ for the incident energy range $35-100$ A\,\textrm{MeV}. We also plot in Fig. \ref{Factor_Z=1} some parametrizations taken from recent works and currently used in transport models 
\cite{LiMachleidt1993,Klakow1993,Schnell1998,Xiangzhou1998,Danielewicz2002}. 
They give rather different results in the Fermi energy domain, showing that $\sigma^{in-medium}_{NN}$ is poorly constrained at present time. All theoretical prescriptions are density-dependent and can give different results when changing density, taken here at a 
fixed value $\rho=1.2\rho_0$. Nevertheless, scanning the expected density 
values $\rho=1-1.5\rho_0$ in this incident energy domain, we have observed for 
$\mathcal{F}$ only small differences of $\pm10\%$, encompassed by the experimental error 
bars. 

From this comparison, we see that the parametrization of the \emph{MSU} group \cite{Danielewicz2002,Coupland2011} is in excellent agreement -within the error bars-  with our experimental findings. The other prescriptions are unable to reproduce the overall trend in the considered energy range, neither in shape nor in magnitude, except Schnell \emph{et al.} \cite{Schnell1998} and Li \emph{et} Machleidt \cite{LiMachleidt1993} at the highest incident energies ($E_{inc} \geq 90-100$ A\,\textrm{MeV}). We can conclude from this part that the medium (density) effects lead to a strong reduction of the $NN$ cross section (by a factor comprised between $2$ and $5$), and that their energy dependence have to be properly accounted in the range $E_{inc}/A=35-100$ \textrm{MeV}. 

\section{Conclusions}

We evaluated nuclear stopping from measured isotropy ratio for protons in 
central collisions for a large body of symmetric systems studied with $INDRA$ 
array. We derived quantitative information on the in-medium transport properties
 in the Fermi energy domain. Firstly, we have found that the stopping is not 
complete above $E_{inc}=30$ A\,\textrm{MeV} whatever the system size. Secondly, we have 
shown that we can get consistent results by scaling the appropriate stopping 
ratio by the characteristic size of the system. We have then established a 
relation between the stopping ratio and the nucleon mean free path in nuclear 
matter. We found $\lambda_{NN}=9.5\pm2~$ \textrm{fm} at $E_{inc}=40$ A\,\textrm{MeV} and $\lambda_{NN}
=4.5 \pm 1$ \textrm{fm} for $E_{inc}=100$ A\,\textrm{MeV}, in agreement with theoretical predictions.
 We also estimated the in-medium effects for the nucleon-nucleon cross 
section by disentangling Pauli blocking effects from higher-order 
correlations due to density (many-body correlations) in nuclear matter. The 
best parametrization is the one provided by Danielewicz \cite{Danielewicz2002}, 
which allows to reproduce the experimental values extracted from this 
analysis. It is interesting to note that this parametrization
has been established in a phenomenological way \cite{Danielewicz2002}. 
We conclude that in-medium effects are quite important since they give a 
significant reduction of the nucleon-nucleon cross section, namely $80\%$ at $E_{inc}=35$ A\,\textrm{MeV} and $50\%$ at $E_{inc}=100$ A\,\textrm{MeV}. This strong energy dependence for the in-medium nucleon-nucleon cross section has to be properly taken into account in any transport model based on Boltzmann equation, where a 2-body collision term is considered. As a perspective, the availability of radioactive beam facilities in the Fermi energy domain could allow to probe more deeply the isopin dependence of the mean free paths, nucleon-nucleon cross-sections but also effective masses. In any case, this could provide valuable information about the isovector properties of the nuclear interaction in dense nuclear matter.


\begin{thebibliography}{99}
%

\bibitem{Onishi1995} A. Ohnishi and J. Randrup, Phys. Rev. Lett. \textbf{75}, 596 (1995).

\bibitem{Aichelin1991} J. Aichelin, Phys. Rep. \textbf{202}, 233 (1991).

\bibitem{Bonasera1994} A. Bonasera \emph{et al.}, Phys. Rep. \textbf{243}, 1 (1994).

\bibitem{Chomaz1994} P. Chomaz, M. Colonna, A. Guarnera and J. Randrup, Phys. Rev. Lett. \textbf{73}, 3512 (1994).

\bibitem{Ono1992} A. Ono, H. Horiuchi, T. Maruyama and A. Ohnishi, Phys. Rev. Lett. \textbf{68}, 2898 (1992).

\bibitem{Gaitanos2005} T. Gaitanos \emph{et al.}, Phys. Lett. B \textbf{609}, 241 (2005).

\bibitem{Kumar2010} S. Kumar, S. Kumar and R.K. Puri, Phys. Rev. C \textbf{81}, 014601 (2010).

\bibitem{Fuchs2006} C. Fuchs and H.H. Wolter, Eur. Phys. J. A \textbf{30}, 5-21 (2006) and refs. therein.

\bibitem{Lattimer2004} J.M. Lattimer and M. Prakash, Science \textbf{304}, 536 (2004). 

\bibitem{Lehaut2010} G. Lehaut \emph{et al.} (INDRA collaboration), Phys. Rev. Lett. \textbf{104}, 232701 (2010).

\bibitem{Durand2006} D. Durand, B. Tamain and E. Suraud, \emph{Nuclear Dynamics 
in the nucleonic regime}, Institute Of Physics, New York (2001) and refs. therein.

\bibitem{Cugnon1996} J. Cugnon, Ann. of Phys., Paris, Vol. \textbf{11} (1996).

\bibitem{Lukasik2005} J. Lukasik \emph{et al.} (INDRA and ALADIN collaboration), Phys. Lett. B \textbf{608}, 223–230 (2005).

\bibitem{LiMachleidt1993} G. Q. Li and R. Machleidt, Phys. Rev. C \textbf{48}, 1702 (1993).

\bibitem{Frick2005} T. Frick, H. Muther, A. Rios, A. Polls and A. Ramos, Phys. Rev. C \textbf{71}, 014313 (2005).

\bibitem{Soma2008} V. Soma and P. Bozek, Phys. Rev. C \textbf{78}, 054003 (2008).

\bibitem{Rios2012} A. Rios and V. Soma, Phys. Rev. Lett. \textbf{108}, 012501 (2012).

\bibitem{Westfall1993} G.D. Westfall \emph{et al.}, Phys. Rev. Lett. \textbf{71}, 1986 (1993). 

\bibitem{Andronic2006} A. Andronic \emph{et al.}, Eur. Phys. J. A \textbf{30}, 
31 (2006) and refs. therein.

\bibitem{Pouthas1995} J. Pouthas \emph{et al.}, Nucl. Inst. and Meth. A \textbf{357}, 418-442 (1995).

\bibitem{Borderie2002} B. Borderie, J. Phys. G \textbf{28}, R217 (2002).

\bibitem{Plagnol1999} E. Plagnol \emph{et al.} (INDRA Collaboration), Phys. Rev.
 C \textbf{61}, 014606 (1999).

\bibitem{Metivier2000} V. Métivier \emph{et al.} (INDRA Collaboration), Nucl. Phys. A \textbf{672}, 357 (2000).

\bibitem{Hudan2003} S. Hudan \emph{et al.} (INDRA Collaboration), Phys. Rev. C \textbf{67}, 064613 (2003).

\bibitem{Lukasik2002} J. Lukasik \emph{et al.} (INDRA and ALADIN Collaborations), Phys. Rev. C \textbf{66}, 064606 (2002).
`
\bibitem{LeFevre2004} A. Le Fevre \emph{et al.} (INDRA and ALADIN Collaborations), Nucl. Phys. A \textbf{735}, 219 (2004).

\bibitem{Zhang2011} G.Q. Zhang \emph{et al.}, Phys. Rev. C \textbf{84}, 034612 (2011).

\bibitem{Charagi1990} S.K. Charagi and S.K. Gupta, Phys. Rev. C \textbf{41}, 1610-1618 (1990).

\bibitem{Renberg1972} P.U. Renberg, D.F. Measday, M. Pepin, P. Schwaller, B. Favier, and C. Richard-Serre, Nucl. Phys. A \textbf{183}, 81-104 (1972).

\bibitem{Nadasen1981} A. Nadasen \emph{et al.}, Phys. Rev. C \textbf{23} 1023-1044  (1981).

\bibitem{Danielewicz1995} P. Danielewicz, Phys. Rev. C \textbf{51} 716 (1995).

\bibitem{Chen2013} B. Chen, F. Sammarruca and C.A. Bertulani, Phys. Rev. C \textbf{87}, 054616 (2013).

\bibitem{Su2013} J. Su and F.-S. Zhang, Phys. Rev. C \textbf{87}, 017602 (2013).

\bibitem{Kikuchi1968} K. Kikuchi and M. Kawai, \emph{Nuclear matter and Nuclear Collisions}, North Holland, New York (1968)

\bibitem{Metropolis1958} N. Metropolis \emph{et al.}, Phys. Rev. \textbf{110}, 204-220 (1958).

\bibitem{Klakow1993} D. Klakow, G. Welke, and W. Bauer, Phys. Rev. C \textbf{48}, 1982-1987 (1993).

\bibitem{Schnell1998} A. Schnell, G. Ropke, U. Lombardo, and H.J. Schulze, Phys. Rev. C \textbf{57}, 806-810 (1998).

\bibitem{Xiangzhou1998} C. Xiangzhou \emph{et al.}, Phys. Rev. C \textbf{58}, 572-575 (1998).

\bibitem{Danielewicz2002} P. Danielewicz, Acta. Phys. Pol. B \textbf{33}, 45 (2002).

\bibitem{Coupland2011} D.D.S. Coupland, W.G. Lynch, M.B. Tsang, P. Danielewicz and Y. Zhang, Phys. Rev. C \textbf{84}, 054603 (2011).


\end{thebibliography}
\end{document}